\documentclass[12pt, letter]{spieman}  % 12pt font required by SPIE;
\usepackage{amsmath,amsfonts,amssymb}
\usepackage{graphicx}
\usepackage{setspace}
\usepackage{tocloft}
\usepackage{lineno}

\usepackage{verbatim} % Added by jlp for the comment comment
\usepackage{xspace}
\usepackage{mathtools}
\usepackage{multirow}
\usepackage{microtype}
\usepackage{cbar}
\usepackage{booktabs}
\usepackage{xcolor,colortbl}
\usepackage{subcaption}
\usepackage{xurl}

\usepackage{caption}
\usepackage{adjustbox}
\usepackage{tikz} % For adding overlays (insets)
\usepackage[table,xcdraw]{xcolor}

\colorlet{rowGreen}{green!15}

\DeclareMathOperator*{\argmin}{arg\,min\xspace}

\title{Synthetic multi-inversion time magnetic resonance images for visualization of subcortical structures}

\author[a,*]{Savannah~P.~Hays}
\author[b]{Lianrui~Zuo}
\author[a]{Anqi~Feng}
\author[b]{Yihao~Liu}
\author[c]{Blake~E.~Dewey}
\author[d]{Jiachen~Zhuo}
\author[c]{Ellen~M.~Mowry}
\author[c]{Scott~D.~Newsome}
\author[a]{Jerry~L.~Prince}
\author[a]{Aaron~Carass}

\affil[a]{Johns Hopkins University, Image Analysis and Communications Laboratory, Dept. of Electrical and Computer Engineering, Baltimore, USA}
\affil[b]{Vanderbilt University, Dept. of Electrical and Computer Engineering, Nashville, USA}
\affil[c]{Johns Hopkins University, Dept. of Neurology, Baltimore, USA}
\affil[d]{University of Maryland School of Medicine, Dept. of Diagnostic Radiology and Nuclear Medicine, Baltimore, USA}

\cftpagenumbersoff{figure}
\cftpagenumbersoff{table} 
\begin{document} 
\maketitle

\begin{abstract}
% 250 word max, 4 sections (purpose, approach, results, and conclusion)
Purpose:
Visualization of subcortical gray matter is essential in neuroscience and clinical practice, particularly for disease understanding and surgical planning.
%While multi-inversion time~(multi-TI) T$_1$-weighted~(T$_1$-w) magnetic resonance~(MR) imaging improves visualization, it is rarely acquired in clinical settings.
\Modified{While multi-inversion time (multi-TI) T$_1$-weighted~(T$_1$-w) magnetic resonance (MR) imaging improves visualization, it is only acquired in specific clinical settings and not available in common public MR datasets.}{1}{MODIFIED}

Approach:
We present SyMTIC (Synthetic Multi-TI Contrasts), a deep learning method that generates synthetic multi-TI images using routinely acquired T$_1$-w, T$_2$-weighted~(T$_2$-w), and FLAIR images.
Our approach combines image translation via deep neural networks with imaging physics to estimate longitudinal relaxation time~($T1$) and proton density~($\rho$) maps.
These maps are then used to compute multi-TI images with arbitrary inversion times.

Results:
SyMTIC was trained using paired MPRAGE and FGATIR images along with T$_2$-w and FLAIR images.
It accurately synthesized multi-TI images from standard clinical inputs, achieving image quality comparable to that from explicitly acquired multi-TI data.
The synthetic images, especially for TI values between 400–800 ms, enhanced visualization of subcortical structures and improved segmentation of thalamic nuclei.

Conclusion:
SyMTIC enables robust generation of high-quality multi-TI images from routine MR contrasts.
\Modified{When paired with the HACA3 algorithm, it generalizes well to varied clinical datasets, including those without FLAIR or T$_2$-w images and unknown parameters, offering a practical solution for improving brain MR image visualization and analysis.}{}{MODIFIED}
%It generalizes well to varied clinical datasets, including those with missing FLAIR images or unknown parameters, offering a practical solution for improving brain MR image visualization and analysis.

\end{abstract}

% Include a list of up to six keywords after the abstract

% Include email contact information for corresponding author
{\noindent \footnotesize\textbf{*}Savannah~P.~Hays,  \linkable{shays6@jhu.edu} }

\begin{spacing}{2}   % use double spacing for rest of manuscript
\section{Introduction}
\begin{figure}
        \centering
        \includegraphics[width = 0.95\linewidth]{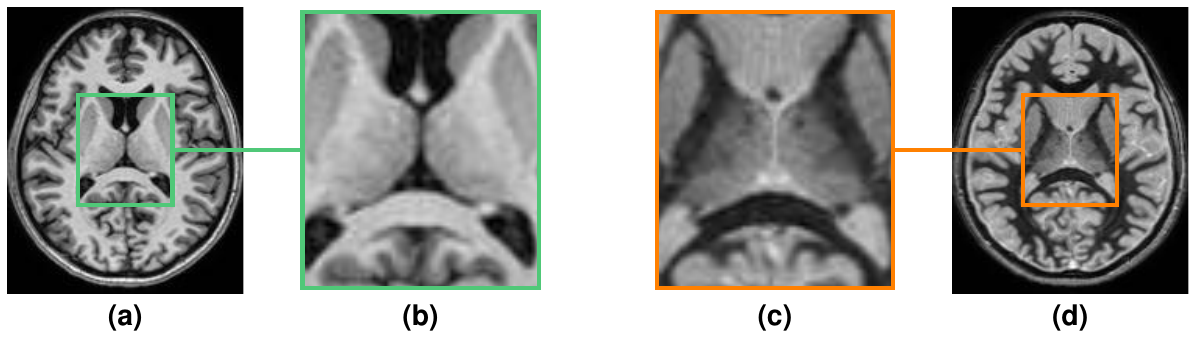}
    \caption{Resulting images from a 3D MPRAGE sequence with two different inversion times~(TIs). \textbf{(a)}~The cerebrospinal fluid nulled Magnetization Prepared Rapid Acquisition with Gradient Echo~(MPRAGE) sequence which is the same sequence with a TI of 1,200ms and \textbf{(d)}~a Fast Gray Matter Acquisition T1 Inversion Recovery~(FGATIR) sequence with a TI of 400ms. Shown in~\textbf{(b)} and~\textbf{(c)} are zoomed versions of~(a) and~(d), respectively, that are focused on the thalamus. The conventional long TI, MPRAGE image shows relatively poor contrast between thalamic nuclei, whereas the short TI, FGATIR images have sufficient contrast to visualize and segment these structures.}
    \label{fig:intro-fgmp}
\end{figure}

Visualization of subcortical structures using magnetic resonance~(MR) imaging
is important for understanding the causes and progression of various 
diseases~\cite{power2015thalamus} and for planning and assessing the efficacy of their 
treatment~\cite{krauss2021technology}. 
Most imaging protocols used to image the thalamus, its nuclei, the basal ganglia, and 
related cortical networks require special pulse sequences such as quantitative susceptibility mapping
(QSM)~\cite{liu2013improved} and fast gray matter acquisition T1 inversion recovery 
(FGATIR)~\cite{sudhyadhom2009high} or specialized hardware such as 7T scanners~\cite{tourdias2014visualization}.
White matter nulled~(WMn) T1-weighted (T$_1$-w)
imaging, such as the FGATIR image shown in Fig.~\ref{fig:intro-fgmp}(d), is often used in surgical planning for deep brain 
stimulation~(DBS)~\cite{krauss2021technology} and is employed in some automatic methods
for segmentation of thalamic nuclei~\cite{su2019thomas, fischl2012freesurfer}.
Despite its increasing usage in clinical protocols, FGATIR is not yet part of many widely used neuroimaging datasets, such as \Modified{OASIS~\cite{LaMontagne} and ADNI~\cite{mueller2005adni}}{1}{MODIFIED}; this gap motivated us to explore synthetic generation of FGATIR images.
\Modified{In this paper, we present a new way to synthesize FGATIR images and to use these to help visualize and delineate thalamic nuclei and related structures using
conventional clinical MR imaging.}{1}{MODIFIED}

The 3D magnetization prepared rapid acquisition with gradient echo~(MPRAGE) image~\cite{mugler1990three} is routinely acquired
in neuroimaging, largely because of its excellent gray matter~(GM)/white 
matter~(WM) contrast as well as its fast acquisition time and high resolution. 
It is conventionally acquired to yield a T$_1$-w contrast 
using a $180^\circ$ inversion pulse and an inversion time~(TI) that will
null the signal from cerebrospinal fluid, as shown in Fig.~\ref{fig:intro-fgmp}(a).
% When DBS is anticipated, it has become common to acquire a FGATIR image as well~\cite{krauss2021technology}. 
The FGATIR pulse sequence is a 3D MPRAGE sequence with a much shorter TI (typically around 400~ms at 3T), which is selected to approximately
null the WM signal.
Although a conventional MPRAGE image shows relatively poor contrast of the thalamic nuclei, striatum, globus pallidus interna, nucleus accumbens, and internal capsule, FGATIR images provide improved contrast relative to MPRAGE for the visualization and/or segmentation of these critical brain structures~\cite{sudhyadhom2009high}.

%
% Varieties of MPRAGE arising from Multi-TI.
%
The difference between the MPRAGE and FGATIR sequences comes
from their different TI values, which lead to significant differences in image contrast.
Acquiring additional MPRAGE images with different TI values is typically not feasible in clinical practice due to time constraints.
However, when at least two MPRAGE images with different TIs (e.g., conventional MPRAGE and FGATIR) are available, they can be used to estimate $T1$ and $\rho$ maps~\cite{brown2014magnetic}.
These tissue parameter maps can then be used with imaging equations to calculate contrast-weighted MR images with arbitrary parameters~\cite{bobman1985cerebralmrisynthesis,sooyeon2022syntheticmri} and multi-TI images.
%
% Imaging equation based synthesis.
%
Jog et al.~\cite{jog2017random} introduced an approach combining machine learning-based synthesis with imaging equations, where estimated pulse sequence parameters were used to modify an atlas image to improve synthesis quality.
However, their method relied on prior tissue segmentation to estimate these parameters, making it less applicable in clinical settings where automated segmentation is not always reliable.
Unlike their approach, our method directly estimates $T1$ and $\rho$ maps from commonly acquired images without requiring segmentation.

Some recent work~\cite{umapathy2022neuroinformatics, moya2021biomedicine,
tohidi2023synthesis} has focused on synthesizing FGATIR images without segmentation.
Umapathy~et~al.~\cite{umapathy2022neuroinformatics} used a single
MPRAGE image to synthesize an FGATIR image using a deep
network, showing better performance in thalamus segmentation on the
synthesized FGATIR than on the acquired MPRAGE.
Moya-S\'{a}ez~et~al.~\cite{moya2021biomedicine} calculated $T1$,
$T2$, and $\rho$ parameter maps from T$_1$-w and T$_2$-w images, training
a deep learning approach with synthetic data to generate an
FGATIR image.
They observed a degradation in
performance of their model when using non skull-stripped images.
Tohidi~et~al.~\cite{tohidi2023synthesis} introduced an additional loss
during network training, using an imaging equation when synthesizing
the $T1$ and $\rho$ parameter maps, reporting superior performance
despite using a small dataset and skull-stripped images.
Skull-stripping has previously been reported to make the synthesis
task easier~\cite{roy2013tmi}; however, it eliminates skull
adjacent regions of interest including the subarachnoid space.
Our previous work~\cite{hays2024spie} synthesized non-skull-stripped
FGATIR images using supervised and unsupervised approaches, with a
3D U-Net showing better performance than using multi-atlas registration~\cite{liu2024jmi}.
However, this approach did not solve for
quantitative parameter maps and was limited to directly synthesizing
an FGATIR image.
More closely related to this work is
Middlebrooks~et~al.~\cite{middlebrooks2023functionalneuro}, which
targeted surgical planning for DBS. They calculated $T1$
maps from the MP2RAGE sequence, which provides two images with
different TIs, and used a simplified imaging equation, which nullified
the $\rho$ component.
As the approach of Middlebrooks~et~al. requires a
MP2RAGE sequence, it has limited applicability as such data is not commonly available in public datasets.

%
% Summary.
%
Building on our previous work~\cite{tohidi2023synthesis}, we present SyMTIC (Synthetic Multi-TI Contrasts), a method that synthesizes $T1$ and $\rho$ parameter maps from three commonly acquired MR images: T$_1$-w, T$_2$-weighted~(T$_2$-w), and fluid-attenuated inversion recovery~(FLAIR).
% SyMTIC integrates deep learning with physics-based modeling, ensuring both interpretability and robustness.
% This approach is broadly applicable to clinical and research datasets, as these MR contrasts are routinely acquired.
% The estimated T$_1$ and PD maps enable the computation of multi-TI images using the appropriate imaging equations, offering a flexible and generalizable solution for visualizing subcortical structures. 
% In our experiments, we show how SyMTIC can handle datasets with missing FLAIR images, varying imaging acquisition parameters, and pathology, all of which are common in clinical imaging.
% Also, SyMTIC retains the skull region to allow downstream analysis of the regions between the brain
% and skull.
Unlike prior approaches, SyMTIC does not rely on specialized acquisitions or skull-stripped data, and it generalizes across domains using harmonization.
The key contributions of this work are:

\begin{itemize}
    %\item We introduce a physics-informed learning framework that enables synthesis of multi-TI images at arbitrary TIs, rather than at a fixed set of values.
    \item We introduce a physics-informed data synthesis model that enables the generation of multi-TI images at arbitrary TIs, rather than at a fixed set of values. Physics-based models are used to synthesize $T1$ and $\rho$ parameter maps, which form the basis for training the deep learning model.
    \item We demonstrate three different training strategies that optimize parameter map accuracy and image synthesis quality.
    \item We retain the full head, including the skull region, to support downstream analyses such as skull-based registration or magnetic resonance-guided focused ultrasound surgerical~(MRgFUS) planning, which most existing methods exclude.
    \item \Modified{By incorporating contrast imputation through harmonization, we show that SyMTIC remains effective even when only a T$_1$-w image is available.}{1}{MODIFIED}
    %By incorporating contrast imputation through harmonization, we show that SyMTIC remains effective even when one of the key input images (FLAIR) is missing.
    \item We evaluate SyMTIC on out-of-domain datasets, demonstrating that the model generalizes well to clinical images acquired with different scanners and protocols.
\end{itemize}
Our model is open source and publicly available from
\url{https://github.com/UponAcceptance}. While the broader capability of generating arbitrary contrasts is a key strength, we focus here on FGATIR due to its practical importance and recognized clinical impact. Contrasts such as those that null the GM-WM boundary are rare in both research and clinical practice, and while capable of generation, this was not the primary focus of our study.

\section{Methods}
\label{s:methods}
\subsection{Overview}
\label{ss:overview}
\begin{figure}[!tb]
   \begin{center}
   \includegraphics[width = 0.9\textwidth]{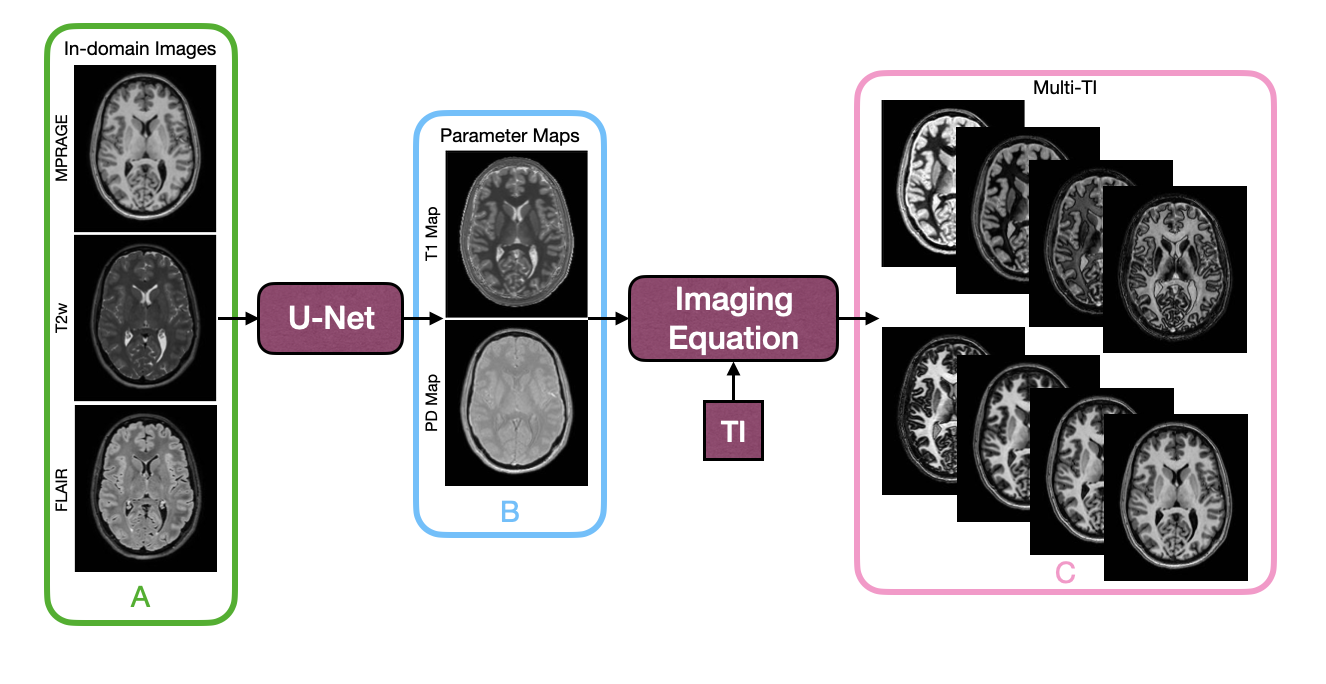}
   \caption{Overview of SyMTIC. \textbf{(A)}~Required in-domain images input to the U-Net model.
   \textbf{(B)}~Synthesis of the $T1$ and $\rho$ parameter maps.
   \textbf{(C)}~Calculation of multi-TI images using the imaging equation and specific TIs.}
   \label{f:model_diagram}
    \end{center}
\end{figure}

Figure~\ref{f:model_diagram} provides an overview of SyMTIC, which can be summarized as consisting of three steps:
in-domain images input to the U-Net model~(Fig.~\ref{f:model_diagram}A),
synthesis of the $T1$ and $\rho$ parameter
maps~(Fig.~\ref{f:model_diagram}B), and application of
imaging equations~(Fig.~\ref{f:model_diagram}C) to generate
multi-TI images.
During training, paired MPRAGE and FGATIR images were used to estimate the ground truth parameter maps.
In the following sections, we discuss our data, preprocessing steps, imaging equations,
synthesis model training, and test time harmonization and imputation.

\subsection{Data}
\begin{figure}[!tb]
    \centering
    \includegraphics[width=0.95\linewidth]{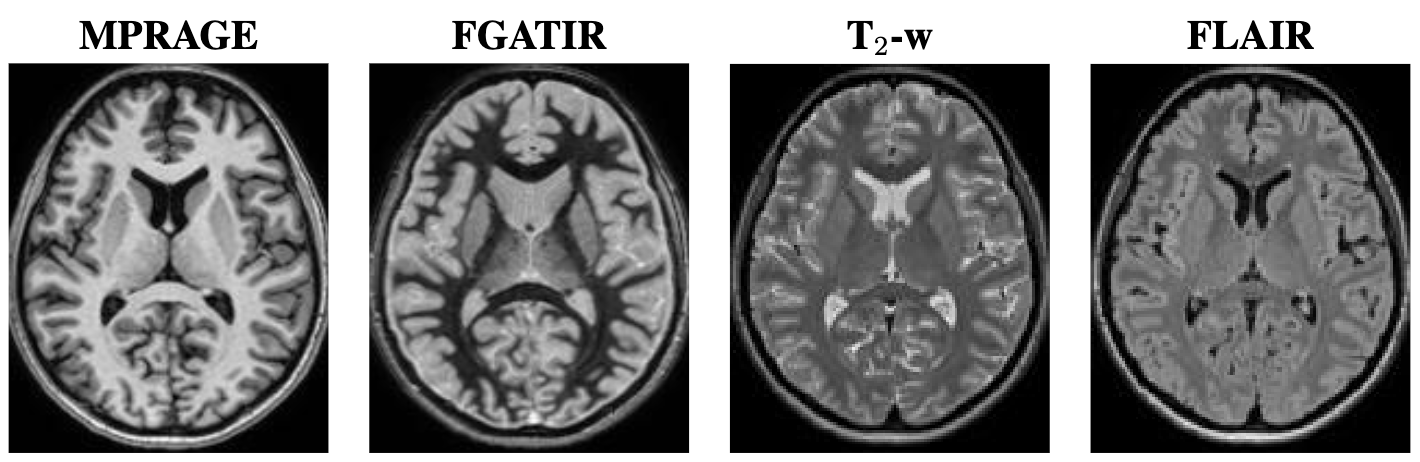}
    \caption{Example images from the same subject, from left to right, are the MPRAGE, FGATIR, T$_2$-w, and FLAIR.}
    \label{f:dataset}
\end{figure}

Our training dataset consists of four brain MR image contrasts that
have been acquired for 23 subjects.
These contrasts include MPRAGE, FGATIR, T$_2$-w, and FLAIR.
Example images of these four contrasts are
shown in Fig.~\ref{f:dataset}.
The MPRAGE and FGATIR images
are acquired with the same imaging parameters (TR = 4,000ms, echo time = 3.37ms, flip angle = $6^\circ$, and magnetic field strength = 3 Telsa) except for their inversion times (TIs), which are TI = 1,400ms for the MPRAGE and TI = 400ms
for the FGATIR.
For training
the networks described in Sec.~\ref{ss:synthesis}, we used a 5-fold cross validation.
We divided our 23
subjects into 14 training subjects, 4 validation subjects, and 5
testing subjects.
For the 5th-fold, only 3 testing subjects were used to allow each subject to be tested once.
The validation set was used to monitor the loss during training and for early stopping.

\subsection{Preprocessing}
Convolutional neural networks~(CNNs) generally perform better when bias field correction and intensity normalization are applied, as these preprocessing steps help mitigate image intensity variations that are unrelated to tissue properties.
We use N4 bias field correction~\cite{tustison2010n4itk}, which is a robust and widely used method, followed by intensity normalization using white matter mean normalization as described in Reinhold et al.~\cite{reinhold2019evaluating}, following the approach of Tohidi et al.~\cite{tohidi2023synthesis}.
For our methodology, preprocessing must be handled carefully for the MPRAGE and FGATIR images due to their direct mathematical relationship.
Since these images are acquired with identical imaging parameters except for their TIs, it is essential to maintain their quantitative relationship to ensure that the imaging equations remain valid.
To perform bias field correction while preserving this relationship, we first estimate the bias field for both the MPRAGE and FGATIR images separately and then compute the geometric mean of these two bias fields.
This single bias field is then applied to correct each image, ensuring that the relative intensity scaling between them remains consistent.
For white matter mean normalization, we first normalize the MPRAGE image and then apply the same normalization factor to the FGATIR image, maintaining their relative intensity relationship.
For all 2D MR acquisitions, we apply a super-resolution~\cite{remedios2024spie} preprocessing step to upsample the images to isotropic 3D resolution. Finally, 3D acquisitions and super-resolved 2D acquisitions are
rigidly registered to the MNI152 atlas using ANTs~\cite{avants2009ants}.

\subsection{Physics-Based Synthesis of Multi-TI Images}
\label{ss:imgeqn}
As described in Hornak~\cite{hornak2010mri}, the intensity $I$ at any voxel $v$ in an inversion recovery sequence can be modeled by the following equation:

\begin{equation}
    I(v) =
\rho(v)\left[1 - 2 \exp{\left(\frac{-\mathrm{TI}}{T1(v)} \right)} + \exp{\left( \frac{-\mathrm{TR}}{T1(v)} \right)} \right],
\label{eqn:imgeqn}
\end{equation}

where $I(v)$ is the image intensity at $v$, $\rho(v)$ is the $\rho$ map
value at $v$, $T1(v)$ is the $T1$ map value at $v$, and $\mathrm{TR}$ is
the repetition time, which is 4,000ms for both our MPRAGE and FGATIR images.
Since the repetition time is fixed and known from the acquisition protocol, and the $\mathrm{TI}$ values for the acquired MPRAGE,~$\mathrm{TI}_{1}$, and FGATIR,~$\mathrm{TI}_{2}$, images are also known, we can solve for the two unknowns, $T1$ and $\rho$ at each voxel.
We rewrite Equation~\ref{eqn:imgeqn} as:
\begin{equation}
    f(x,y;a,b) = x \left[1 - 2\exp{\left(\frac{-a\phantom{-}}{y}\right)} + \exp{\left(\frac{-b\phantom{-}}{y}\right)} \right],
\label{eqn:fxy_eqn}
\end{equation}
where $x$ and $y$ correspond to $\rho$ and $T1$, respectively.
We then solve the following optimization problem at each voxel $v$ using a least squares formulation:
\begin{eqnarray}
%
    %need to fix the argmin and R dimensions
    (\rho(v), T1(v)) & = & \argmin_{(x,y) \in \mathbb{R}^2} \left( \left(I_1(v) - f(x, y; \mathrm{TI}_{1},
\mathrm{TR})\right)^2 \right. \nonumber\\
&& \left. \qquad + \left(I_2(v) - f(x, y; \mathrm{TI}_{2}, \mathrm{TR})\right)^2
\right) ,
\label{eqn:mpdmt1}
\end{eqnarray}
where $I_1$ is the acquired MPRAGE image, $I_2$ is the acquired FGATIR image, and $\mathrm{TI}_{1}$ and $\mathrm{TI}_{2}$ are their corresponding inversion times.
Although Eq.~\ref{eqn:mpdmt1} assumes signed signal values, most MR images are reconstructed and stored as magnitude images.
In particular, the FGATIR sequence is acquired at a very short TI (e.g., TI = 400 ms), where the longitudinal magnetization for many tissues is still negative prior to readout, especially those in and around the deep brain structures.
To correctly solve for ${\rho}$ and ${T1}$ using Eq.~\ref{eqn:mpdmt1}, we therefore negate the FGATIR image values prior to optimization to reflect the signed nature of the signal predicted by the inversion recovery equation.
This detail is critical for accurate parameter estimation; however, for rendering synthetic multi-TI images from Eq.~\ref{eqn:imgeqn}, we apply the absolute value to match the magnitude format of clinically acquired MR images.

% $M_{PD}(v)$ is the PD map value at $v$, $M_{T1}(v)$ is the T$_1$
% map value at $v$, $\mathbb{R}_0$ is the set of non-negative real
% numbers, $\times$ is the Cartesian product, $I_1$ is the MPRAGE image,
% $I_2$ is the FGATIR image, $T_{I1}$ is the TI for the MPRAGE image,
% $T_{I2}$ is the TI for the FGATIR image, and $T_R$ is the TR.

\begin{figure}[!tb]
    \centering
    \includegraphics[width=0.9\linewidth]{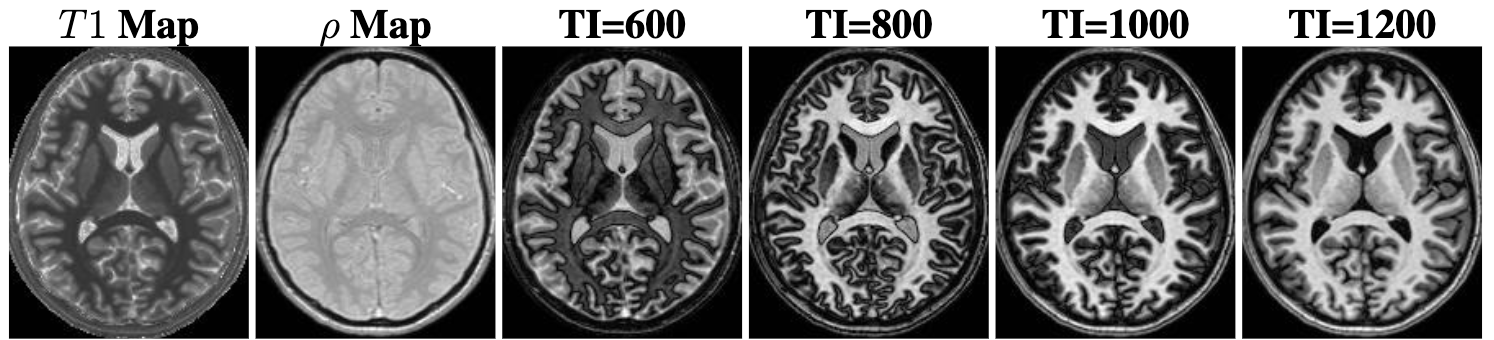}
    \caption{Example of parameter maps and multi-TI images computed from the MPRAGE and FGATIR images shown in Fig.~\ref{f:dataset}.
   \label{f:dataset_multiTI} 
    }  %note label inside caption
\end{figure}

After computing $\rho$ and $T1$ for all voxels,
we can use Eq.~\ref{eqn:imgeqn} to calculate an image at any desired
TI.
We refer to the collection of such images, computed at different TI values, as multi-TI images.
Figure~\ref{f:dataset_multiTI} shows examples of the estimated $T1$ and $\rho$ parameter maps and some computed multi-TI images from the MPRAGE and FGATIR images in Fig.~\ref{f:dataset}.

\subsection{Synthesis Model}
\label{ss:synthesis}
%
% Why we need image synthesis and describe the network.
%
%As noted previously, the FGATIR image is rarely acquired thus it is
%not feasible for us to depend on its existence for generation of T$_1$ and PD maps, which enable the creation of the multi-TI images.
To enable generation of multi-TI images when an FGATIR image is not acquired, we designed a synthesis network using a 2D U-Net~\cite{ronneberger2015miccai} network to
take commonly acquired MR images and synthesize $T1$ and $\rho$ maps, from which multi-TI images can be computed using Eq.~\ref{eqn:imgeqn}.
We trained the network by concatenating corresponding 2D slices
from the MPRAGE, T$_2$-w, and FLAIR images in any orientation into a
three-channel input tensor.
The outputs of this network are $T1$ and $\rho$ maps, which are available for training purposes from our cohort using the
estimation step in Sec.~\ref{ss:imgeqn}.
We use this 2D network to
build a 3D volume by application of the network in all three cardinal
orientations.
We fuse these three outputs by taking their median, yielding 3D $T1$ and $\rho$ maps.
In the encoder portion of our network, we
use max pooling with a $2 \times 2$ kernel and a stride of $2$ to
reduce spatial dimensions.
In the decoder, feature maps are
upsampled and concatenated with corresponding encoded features,
creating skip connections that help preserve spatial context at every
downsampling resolution.
Both the encoder and decoder employ 2D
convolutions followed by instance normalization and LeakyReLU
activation to stabilize training and introduce non-linearity.
The
network's output is processed by a sigmoid function and then scaled by
$4,000$, a modification we found to expedite convergence and allow
parameter maps to reach realistic values more effectively than using a
traditional ReLU activation.
We optimized the network using the Adam optimizer with a learning rate of $\mathrm{10^{-6}}$.

\begin{figure}[!tb]
   \begin{center}
   \includegraphics[width = 0.8\textwidth]{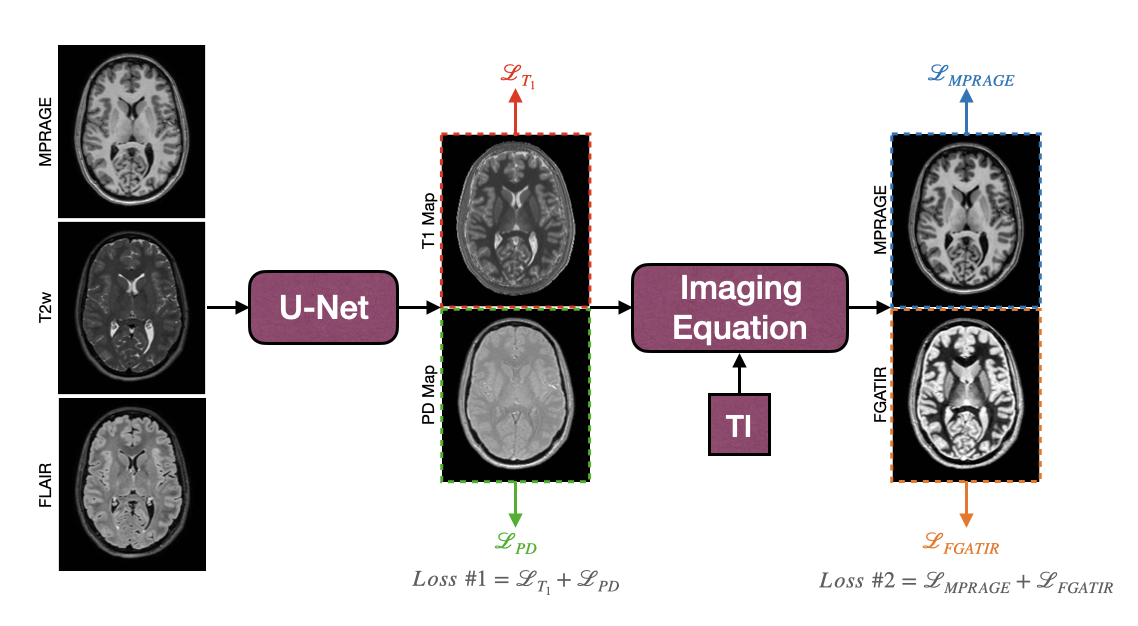}

   \caption{Overview of the different training losses. Loss~\#1 is calculated by the summation of the L1 losses on the predicted and ground truth $T1$ and $\rho$ maps. Loss~\#2 is calculated by the summation of the L1 losses on the predicted and ground truth MPRAGE and FGATIR images.
   \label{f:training_losses}
    }  %note label inside caption
    \end{center}
\end{figure}

We compared using two different losses for training the network, as illustrated in Fig.~\ref{f:training_losses}.
Model~\#1 uses Loss~\#1, which applies the L1 error to the output $T1$ and $\rho$ maps as compared to those computed by Eq.~\ref{eqn:mpdmt1} equivalent to the approach
proposed by Moya-S\'{a}ez~et~al.~\cite{moya2021biomedicine}.
Model~\#2 uses Loss~\#2, which applies the L1 error on the output FGATIR (TI = 400ms) and MPRAGE (TI = 1400ms) images computed from the derived $T1$ and $\rho$ maps similar to that used in
Tohidi~et~al.~\cite{tohidi2023synthesis}.
We can combine Loss~\#1 and Loss~\#2 to form Model~\#3.
In all models, Loss~\#1 is equally weighted between the $T1$ and $\rho$ maps and Loss~\#2 weights the FGATIR image $\times 2$ over the MPRAGE.

\subsection{Test time harmonization and imputation}
\label{ss:testtime}
Our synthetic $T1$ and $\rho$ maps are quantitative, making them sensitive to domain shifts in input images.
Therefore, when applying our model to out-of-domain datasets or datasets with missing contrasts, we employ the HACA3~\cite{zuo2022haca3} MR harmonization algorithm as a preprocessing step after registration to ensure that the input images resemble those used during training.
HACA3 is an open
source\footnote{\url{https://github.com/lianruizuo/haca3}}
unsupervised image harmonization approach for structural MR neuroimages.
It does not require paired subject data for training.
It uses an
encoder-decoder structure to learn latent representations of anatomy,
acquisition contrast, and image quality, enabling harmonization by
combining these representations.
HACA3 has the flexibility to handle single or multiple contrast MR images when harmonizing
and imputing missing MR images when needed.
\Modified{Recent work by Lu et al.~\cite{yuan2025imagingneruoscience} provides a comprehensive evaluation of image-based and statistical MR image harmonization techniques, and demonstrates that HACA3 produces the most reliable and consistent results in comparison to other methods.}{1}{MODIFIED}
When using HACA3, our target image contrasts are the T$_1$-w MPRAGE,
T$_2$-w, and FLAIR images that we used to training our synthesis
model.
The encoded target images are available on our Github page:~\url{https://github.com/UponAcceptance}.
\Modified{When the three images required for SyMTIC are out of domain or missing, we use either the harmonized or synthetic images produced by HACA3 as inputs to SyMTIC.}{1}{MODIFIED}

\section{Experiments and Results}
\subsection{Comparison of training objectives for SyMTIC}
\begin{figure}[!tb]
    \centering
    \includegraphics[width=0.9\linewidth]{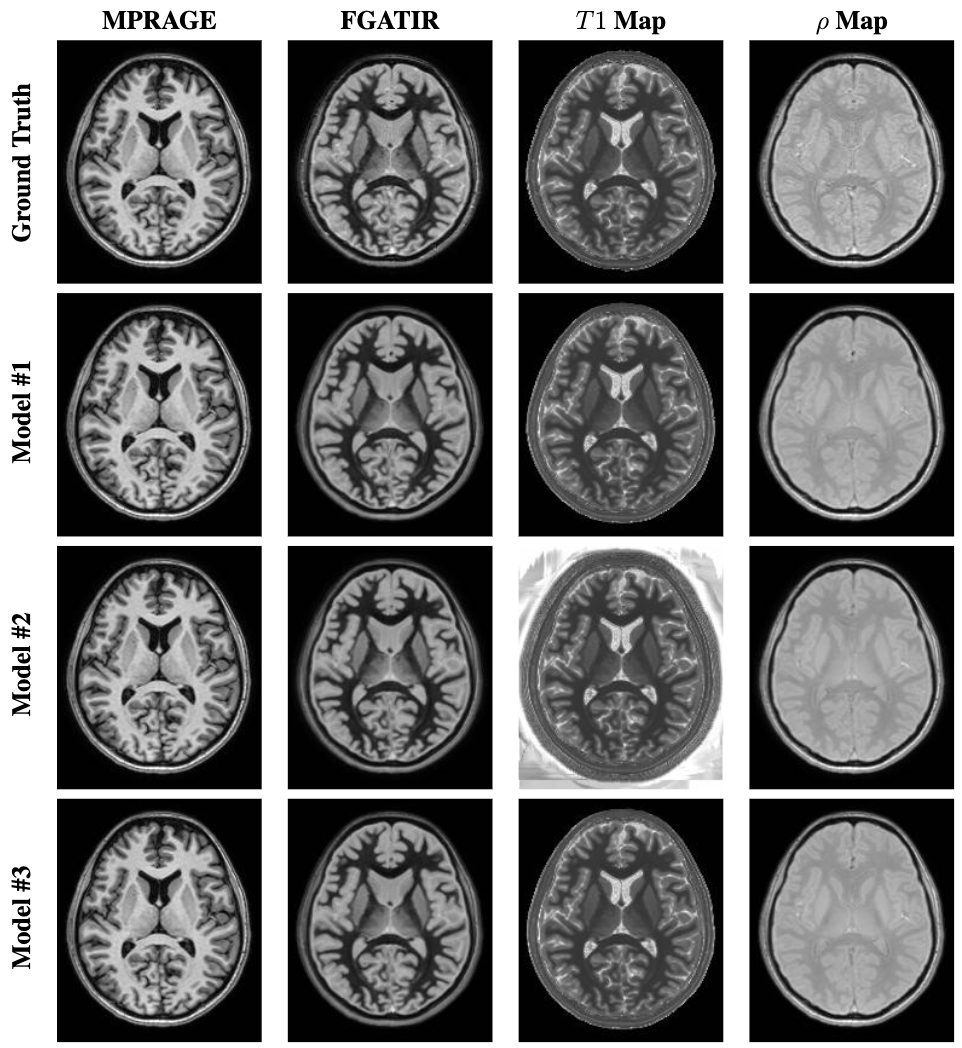}
   \caption{Ground truth images of a testing subject with the prediction images using the proposed SyMTIC models.}
   \label{fig:pred_maps} 
\end{figure}

\begin{figure}[!tb]
    \centering
    \includegraphics[width=0.95\linewidth]{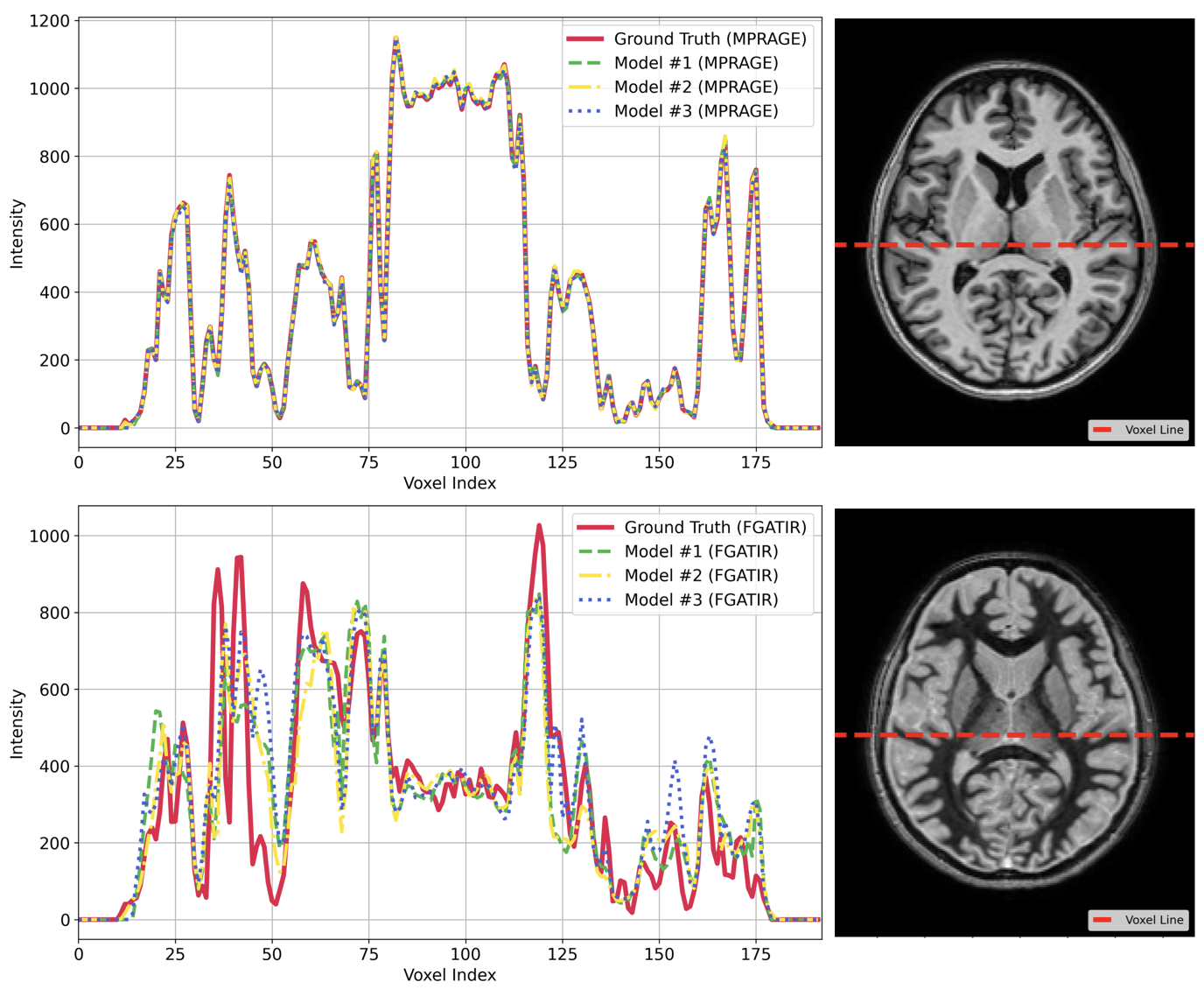}
    \caption{Intensity values through each voxel across the thalamic cross-section shown on the right in the MPRAGE~(top) and FGATIR~(bottom). The plots compare the Ground Truth (GT) intensity values against predictions from each model. The GT line (red) and prediction lines highlight areas of similarity and deviation in voxel intensities across the horizontal axis of the selected slice.}
    \label{fig:voxel_crosssec}
\end{figure}

% or "Loss function comparison for accurate T1 and PD map generation"
First, we tested our models on the in-domain test set ($N=23$).
An illustrative example of the predicted $T1$ and $\rho$ parameter maps, along with the corresponding synthetic MPRAGE and FGATIR images, is shown in Fig.~\ref{fig:pred_maps} with the corresponding ground truth images (the acquired images and computed parameter maps).
Figure~\ref{fig:voxel_crosssec} features a thalamic cross-sectional intensity profile of the MPRAGE and FGATIR ground truth and prediction images. This provides a visual means of evaluating the alignment and fidelity of the predicted images relative to the ground truth.

Table~\ref{tab:indomain_results} reports the PSNR and SSIM for each image synthesized from each model.
To ensure the evaluation focuses on meaningful image content, we excluded background voxels from the error calculations.
Among the three models, we observed similar performances but Model~\#2 give us the highest quantitative results for the two synthetic images and two parameter maps computed by SyMTIC.
In Table~\ref{tab:indomain_results}, the paired Wilcoxon signed-rank test with Bonferroni correction was used to determine statistical significance between the three models and the reference images. Model~\#2 showed statistically significant improvements (p $<0.001$) over both Model~\#1 and Model~\#3 for PSNR and SSIM in MPRAGE and $T1$ maps and only SSIM in FGATIR. 
For the $\rho$ map, while Model~\#1 demonstrated lower variability, Model~\#2 achieved the highest mean SSIM, which was statistically better~(p $<0.05$) than Model~\#1. 
These results support the overall superiority of Model~\#2 for synthesizing high-fidelity parameter maps and images.

\begin{table}[!tb]
    \centering
    \caption{Calculated PSNR and SSIM values on the synthetic images and parameter maps using the proposed SyMTIC models ($N=23$). Significant differences between models are indicated using a paired Wilcoxon test with Bonferroni correction (\(^*\): significantly higher~(p $< 0.001$) compared to Models \#1 and \#3, \(^\dagger\): significantly higher~(p $<0.05$) compared to Model \#1). The highest value for each image is indicated in \textbf{bold}.}
    \begin{tabular}{c c c c c c}
        \toprule
        \textbf{Model} & \textbf{Loss \#1} & \textbf{Loss \#2} & \textbf{Image} & \textbf{PSNR} & \textbf{SSIM} \\
        \cmidrule(lr){1-6}
        \multirow{4}{*}{Model \#1} & \multirow{4}{*}{\checkmark} & \multirow{4}{*}{--} & MPRAGE & 40.33 $\pm$ 3.11 & 0.9512 $\pm$ 0.0133 \\
        &  &  &\cellcolor{rowGreen} FGATIR &\cellcolor{rowGreen} 26.91 $\pm$ 1.34 &\cellcolor{rowGreen} 0.7334 $\pm$ 0.0472 \\
        &  &  & $T1$ Map & 22.10 $\pm$ 0.99 & 0.7804 $\pm$ 0.0247 \\
        &  &  &\cellcolor{rowGreen} $\rho$ Map &\cellcolor{rowGreen} 29.21 $\pm$ 0.84 &\cellcolor{rowGreen} 0.8381 $\pm$ 0.0298 \\
        \cmidrule(lr){1-6}
        \multirow{4}{*}{Model \#2} & \multirow{4}{*}{--} & \multirow{4}{*}{\checkmark} & MPRAGE & \textbf{45.70} $\pm$ 5.67\(^*\) & \textbf{0.9970} $\pm$ 0.0029\(^*\) \\
        &  &  &\cellcolor{rowGreen} FGATIR &\cellcolor{rowGreen} \textbf{27.60} $\pm$ 2.27 &\cellcolor{rowGreen} \textbf{0.7906} $\pm$ 0.0535\(^*\) \\
        &  &  & $T1$ Map & \textbf{23.75} $\pm$ 1.73\(^*\) & \textbf{0.8088} $\pm$ 0.0319\(^*\) \\
        &  &  &\cellcolor{rowGreen} $\rho$ Map &\cellcolor{rowGreen} \textbf{29.82} $\pm$ 2.14 &\cellcolor{rowGreen} \textbf{0.8516} $\pm$ 0.0365\(^\dagger\) \\
        \cmidrule(lr){1-6}
        \multirow{4}{*}{Model \#3} & \multirow{4}{*}{\checkmark} & \multirow{4}{*}{\checkmark} & MPRAGE & 42.35 $\pm$ 3.73 & 0.9564 $\pm$ 0.0127 \\
        &  &  &\cellcolor{rowGreen} FGATIR &\cellcolor{rowGreen} 26.99 $\pm$ 1.43 &\cellcolor{rowGreen} 0.7408 $\pm$ 0.0492 \\
        &  &  & $T1$ Map & 22.31 $\pm$ 0.97 & 0.7906 $\pm$ 0.0228 \\
        &  &  &\cellcolor{rowGreen} $\rho$ Map &\cellcolor{rowGreen} 29.25 $\pm$ 1.39 &\cellcolor{rowGreen} 0.8437 $\pm$ 0.0305 \\
        \bottomrule
        \\[-0.3em]
    \end{tabular}
    \label{tab:indomain_results}
\end{table}

\subsection{Multi-TI synthesis for in-domain testing dataset}
\begin{figure}[!tb]
    \centering
    \includegraphics[width=0.9\linewidth]{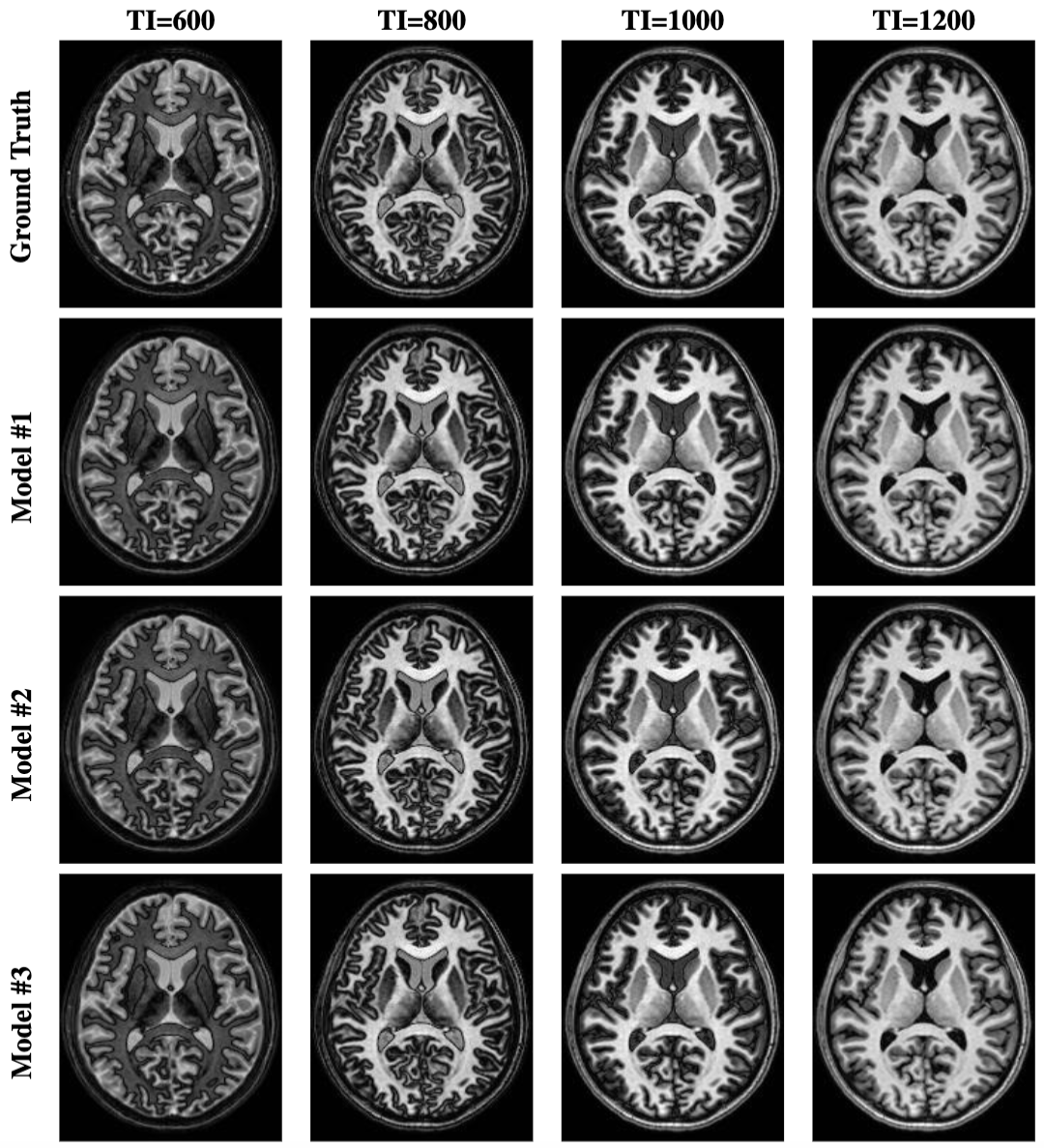}
   \caption{Synthetic multi-TI images of a test subject calculated using the predicted parameter maps from the SyMTIC models in Fig.~\ref{fig:pred_maps}.}
   \label{fig:mtbi_multiti} 
\end{figure}

\begin{figure}[!pt]
    \centering
    \includegraphics[width=0.9\linewidth]{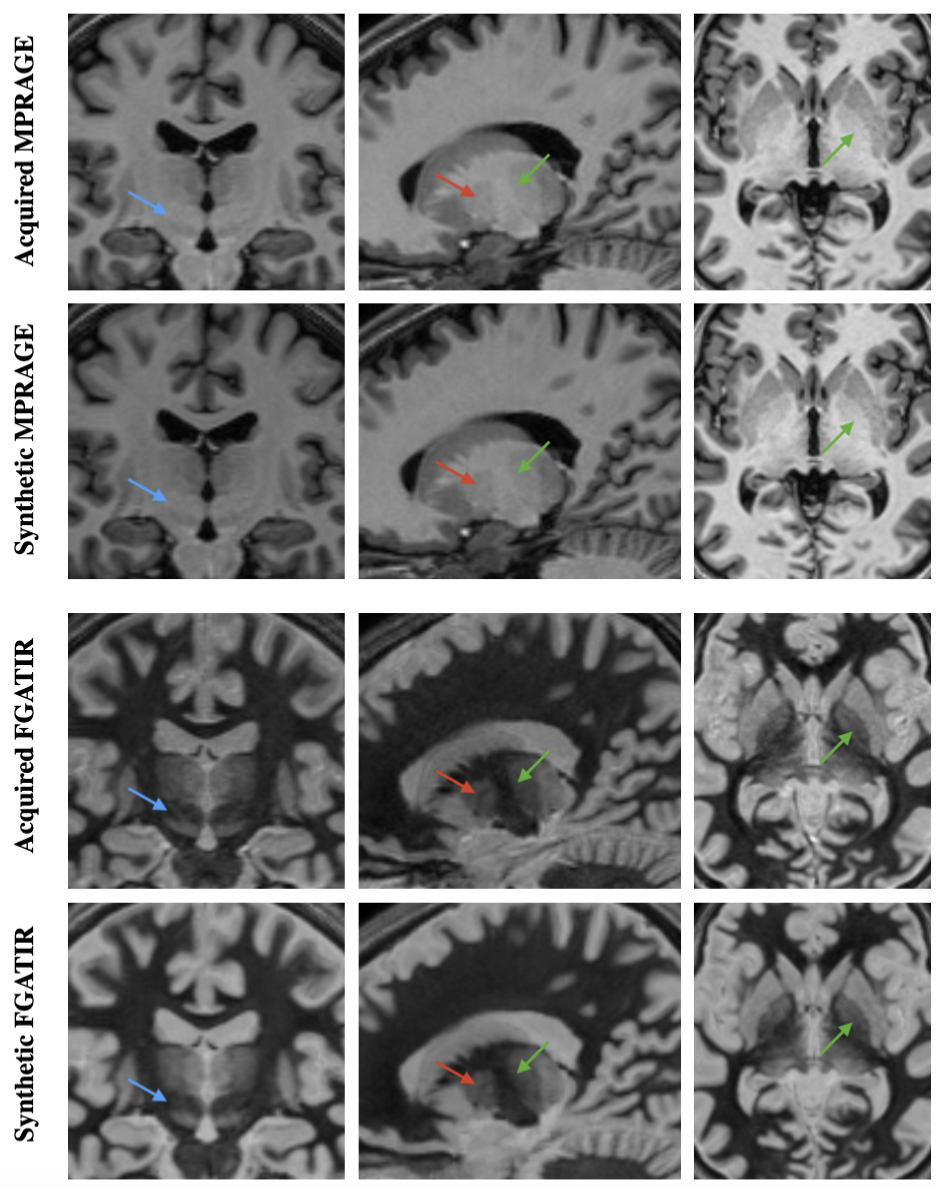}
\caption{Acquired MPRAGE, synthetic MPRAGE, acquired FGATIR, and synthetic FGATIR images (top to bottom). Arrows point to DBS targets that are less visible in MPRAGE images: blue for subthalamic nuclei, red for thalamic nuclei, and green for the internal lamina.}
\label{fig:dbs-targets}
\end{figure}

Figure~\ref{fig:mtbi_multiti} shows examples of multi-TI images computed from the $T1$ and $\rho$ maps corresponding to the ground truth and the three SyMTIC models.
Figure~\ref{fig:dbs-targets} provides a visual comparison of an acquired and synthetic MPRAGE and FGATIR images in the axial, sagittal, and coronal planes.
We highlight DBS targets in these images, demonstrating that these structures are clearly visible in both the acquired and synthetic FGATIR images.
This qualitative comparison supports the potential of our model to enhance DBS planning through synthetic FGATIR image generation.

\subsection{Thalamus segmentation using THOMAS}
\label{sec:thalamusseg}
\begin{figure}[!tb]
    \centering
    \includegraphics[width=0.95\linewidth]{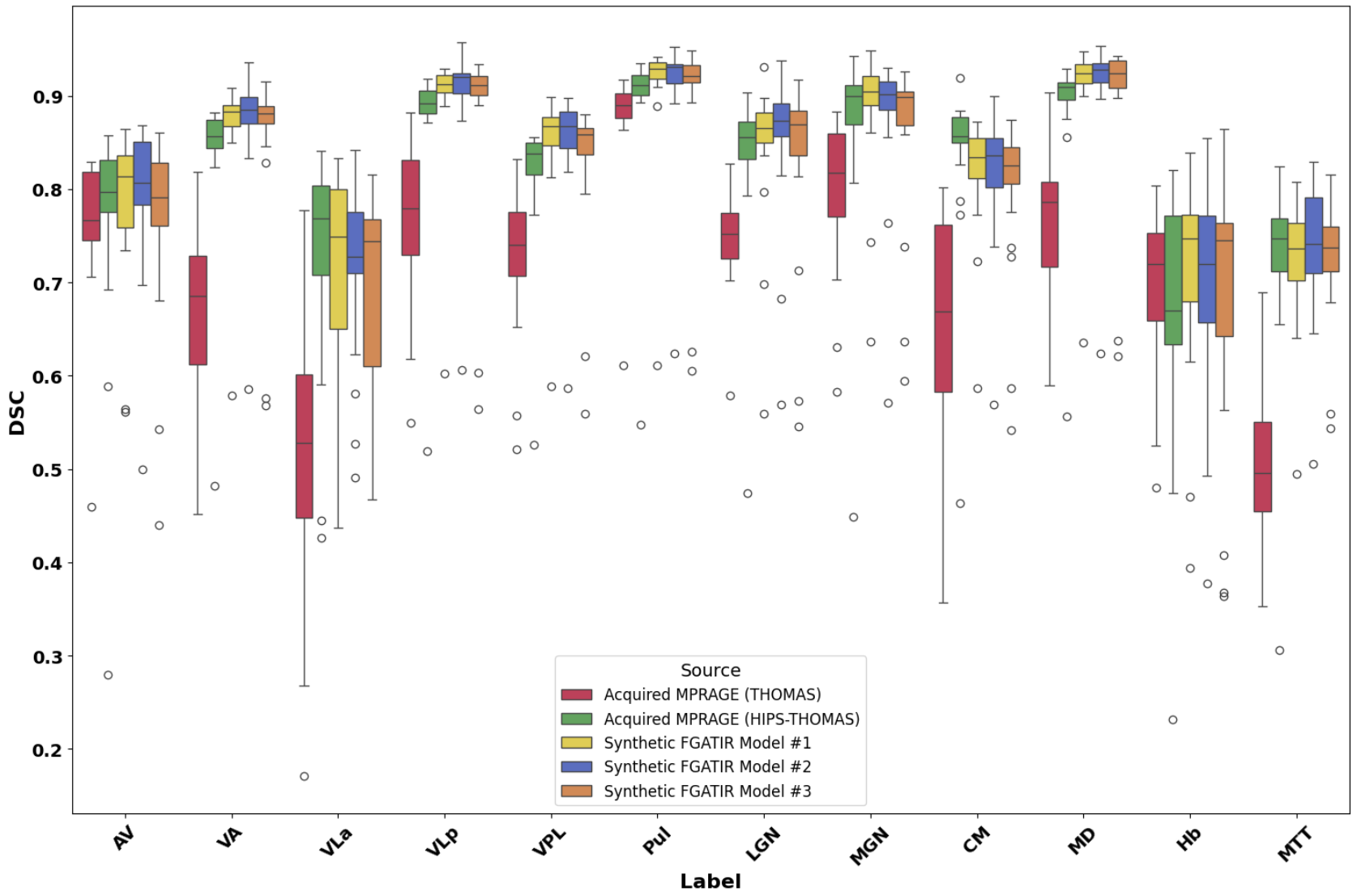}
    \caption{Per-label DSC for thalamic nuclei segmentation on each input image. The 12 labels are the anteroventral~(AV), ventral anterior~(VA), ventral lateral anterior~(VLa), ventral lateral posterior~(VLp), ventral posterolateral~(VPL), pulvinar~(Pul), lateral geniculate~(LGN), medial geniculate~(MGN), centromedian~(CM), mediodorsal~(MD), habenula~(Hb), and mammillothalamic tract~(MTT).}
    \label{fig:dsc_labels}
\end{figure}

It has been reported by Su et al.,~\cite{su2019thomas} that THOMAS achieves the highest thalamus segmentation when using an FGATIR image.
\Modified{HIPS-THOMAS is an enhanced version of the THOMAS thalamus segmentation algorithm.
HIPS-THOMAS uses piecewise polynomial fitting to transform the intensities of a MPRAGE image to match the histogram of a target FGATIR image.
HIPS-THOMAS then performs the standard THOMAS algorithm on the synthetic FGATIR image.}{1(d)}{MODIFIED}
The goal of this experiment is to evaluate whether our synthetic FGATIR images could serve as a viable alternative to acquired FGATIR images for thalamus segmentation.
By comparing THOMAS and HIPS-THOMAS segmentation results across different image types, we aimed to assess whether our synthetic FGATIR images can bridge the gap between standard clinical imaging protocols and the enhanced contrast required for accurate thalamic delineation. 
This validation is particularly important for extending automated thalamic segmentation methods to datasets where FGATIR images are not available, ultimately broadening the clinical utility of standard MR imaging acquisitions.
Since we do not have thalamus labels for our dataset, we used the THOMAS result on the acquired FGATIR image as the ground truth.
We compared with THOMAS and HIPS-THOMAS~\cite{vidal2024thomaships} on the acquired MPRAGE along with the THOMAS result on the synthetic FGATIR images from the three models.
In Table~\ref{table:thomas_dsc}, a summary of the Dice coefficients~(DSC) for thalamus segmentation are reported. Statistical significance using paired Wilcoxon signed-rank tests with Bonferroni correction was calculated. All models achieved a significantly higher (p $<0.001$) DSC compared with using the acquired MPRAGE for FGATIR-based segmentation.
Figure~\ref{fig:dsc_labels} shows the individual nuclei DSC coefficients.
Running THOMAS on the synthetic FGATIR from our Model~\#2 yields a comparable result to running HIPS-THOMAS on an acquired MPRAGE image, which validates our synthetic images.
% The Dice coefficients~(DSC) are reported in Table~\ref{table:thomas_dsc}.
% Our results indicate that THOMAS segments the thalamus more accurately using any of the synthetic FGATIR images than the acquired MPRAGE.
% This improvement is statistically significant across all models (Wilcoxon signed-rank test, p $<0.001$ for all comparisons).

% \begin{table}[!tb]
% \centering
% \caption{The entire thalamus DSC and mean DSC of the 12 classes for different input images to THOMAS.}
% \begin{tabular}{l cc cc}
% \toprule
% \textbf{Input Image} && \textbf{DSC for Entire Thalamus} && \textbf{Mean DSC of 12 Classes} \\
% \cmidrule{1-5}
% \rowcolor{rowGreen}
% Acquired MPRAGE && 0.820 $\pm$ 0.059 && 0.708 $\pm$ 0.063 \\
% %
% % \rowcolor{rowGreen}
% Synthetic FGATIR Model \#1 && 0.894 $\pm$ 0.066 && 0.826 $\pm$ 0.064 \\
% %
% \rowcolor{rowGreen}
% Synthetic FGATIR Model \#2 && \textbf{0.896 $\pm$ 0.067} && \textbf{0.828 $\pm$ 0.067} \\
% % \rowcolor{rowGreen}
% Synthetic FGATIR Model \#3 && 0.878 $\pm$ 0.059 && 0.810 $\pm$ 0.096 \\
% %

% \bottomrule
% \\[-0.3em]
% \end{tabular}
% \label{table:thomas_dsc}
% \end{table}

\begin{table}[!tb]
\setlength{\tabcolsep}{2pt}
\centering
\caption{The entire thalamus DSC and mean DSC of the 12 classes for different input images to THOMAS.
The ground truth segmentation was generated by processing the acquired FGATIR image with THOMAS. 
Synthetic FGATIR images were produced with three different models and then segmented using THOMAS. 
Acquired MPRAGE images were segmented using both THOMAS and HIPS-THOMAS for comparison.
Results indicate that synthetic FGATIR images from all models and acquired MPRAGE (HIPS-THOMAS) yield statistically significantly higher DSC values relative to acquired MPRAGE (THOMAS) (Wilcoxon signed-rank test, p $<0.001$ for all comparisons, denoted by \(^{*}\)).
The highest value in each column is indicated in \textbf{bold}.
}
\begin{tabular}{l cc cc}
\toprule
\textbf{Input Image} && \textbf{DSC for Entire Thalamus} && \textbf{Mean DSC of 12 Classes} \\
\cmidrule{1-5}
\rowcolor{rowGreen}
Acquired MPRAGE (THOMAS) && 0.820 $\pm$ 0.059 && 0.708 $\pm$ 0.063 \\
Acquired MPRAGE (HIPS-THOMAS) && 0.874 $\pm$ 0.081\(^*\) && 0.811 $\pm$ 0.090\(^*\) \\
\rowcolor{rowGreen}
Synthetic FGATIR Model \#1 && 0.894 $\pm$ 0.066\(^*\) && 0.826 $\pm$ 0.064\(^*\) \\
Synthetic FGATIR Model \#2 && \textbf{0.896 $\pm$ 0.067}\(^*\) && \textbf{0.828 $\pm$ 0.067}\(^*\) \\
\rowcolor{rowGreen}
Synthetic FGATIR Model \#3 && 0.878 $\pm$ 0.059\(^*\) && 0.810 $\pm$ 0.096\(^*\) \\

\bottomrule
\\[-0.3em]
\end{tabular}
\label{table:thomas_dsc}
\end{table}

% \subsection{Missing FLAIR image}
\subsection{\Modified{Out-of-domain testing}{1(a,c)}{MODIFIED}}
\label{sec:missingmr}
\begin{figure}[!tb]
    \centering
    \includegraphics[width=0.75\linewidth]{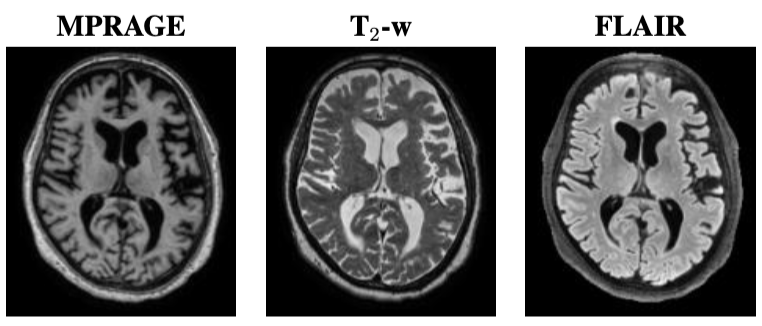}
   \caption{An acquired MPRAGE image, acquired T$_2$-w image, and synthetic FLAIR image from the MMTI dataset. The synthetic FLAIR was imputed using HACA3 and the other two acquired images.}
   \label{fig:mmti_data} 
\end{figure}

\begin{figure}[!tb]
    \centering
    \includegraphics[width=0.9\linewidth]{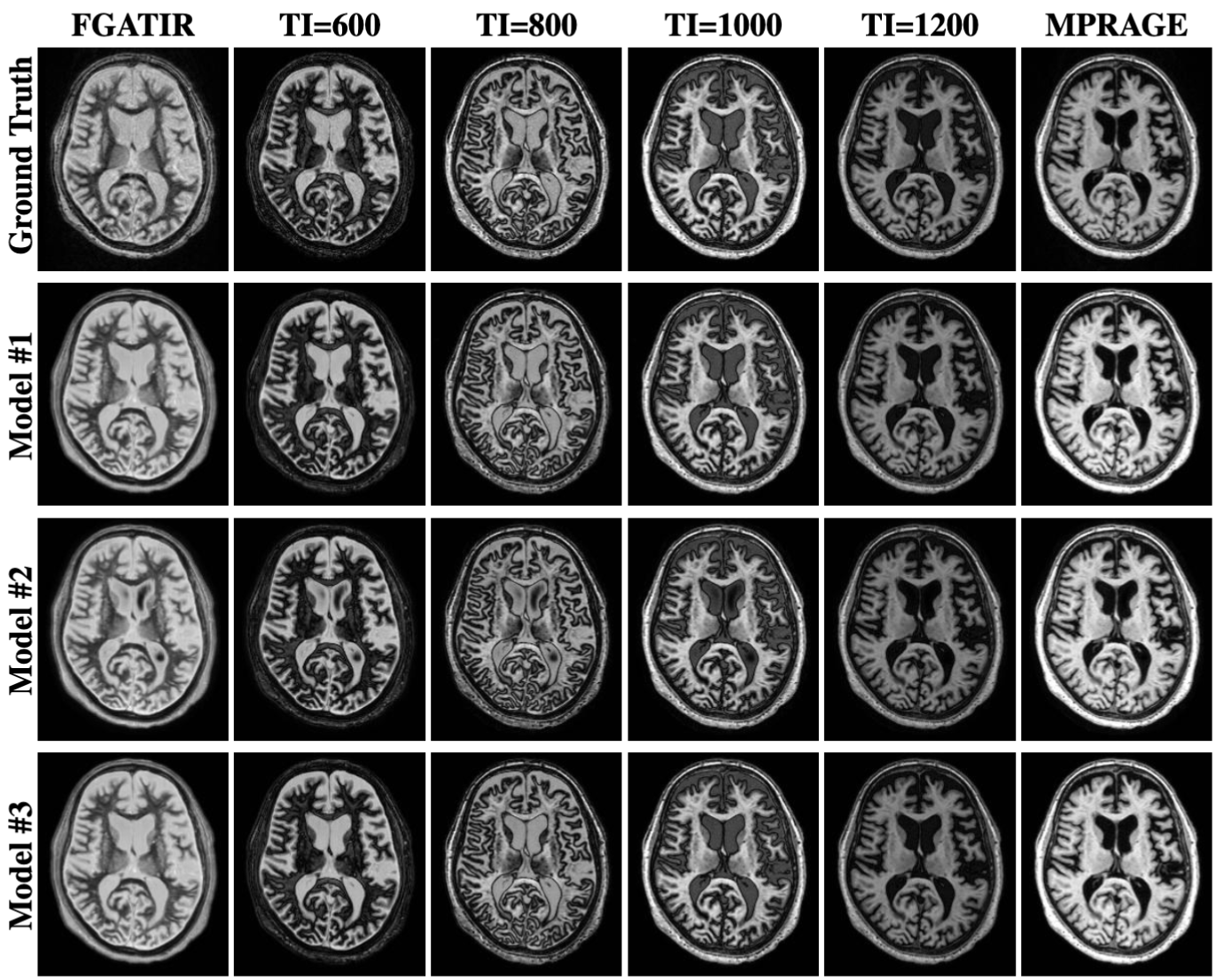}
   \caption{Calculated multi-TI images from the MMTI dataset using Models~\#1-3.}
   \label{fig:mmti_multiTI}
\end{figure}

\Modified{We tested Models~\#1--3 on an out-of-domain dataset ($N=35$) that does not have FLAIR images.}{1(c)}{MODIFIED}
We refer to this dataset as MMTI.
% This dataset  has MPRAGE and FGATIR images so it allowed us to evaluate our model performance on a large set of unseen data.
To address the missing FLAIR image contrast, we used HACA3 to synthesize a FLAIR image using the acquired MPRAGE and T$_2$-w images.
An acquired MPRAGE image, acquired T$_2$-w image, and synthetic FLAIR image from the MMTI dataset are shown in Fig.~\ref{fig:mmti_data}.
From the predicted parameter maps, we calculated the multi-TI images shown in Fig.~\ref{fig:mmti_multiTI}.
This experiment shows that SyMTIC can handle datasets with missing FLAIR images by utilizing HACA3 for imputation.

\Added{To assess HACA3 performance as a preprocessing stage for SyMTIC, we performed four comparisons using SyMTIC Model \#2.
We used HACA3 to impute FLAIR and T$_2$-w images assuming only a T$_1$-w image was acquired, impute only FLAIR assuming only T$_2$-w and T$_1$-w images were acquired, and harmonization of the acquired images in each of the previous scenarios. For a quantitative comparison, we performed thalamus segmentation as done in Section~\ref{sec:thalamusseg}.
The DSC in Table~\ref{table:mmti_thomas_dsc} is computed between the image defined and the ground truth established by the THOMAS algorithm using the acquired FGATIR image. 
All results from the different HACA3 uses lead to statistically improved results in comparison to using THOMAS on an acquired MPRAGE and comparable results to using HIPS-THOMAS on an acquired MPRAGE. 
% This demonstrates the reliably of our method and HACA3 to not only synthesize FGATIR images, but a range multi-TI images.
}{1(c)}{ADDED}
\begin{table}[!h]
\setlength{\tabcolsep}{2pt}
\centering
\caption{The whole thalamus DSC and mean DSC of the 12 classes for different input images.
The ground truth segmentation was obtained by processing the acquired FGATIR image with THOMAS.
Synthetic FGATIR images were generated using SyMTIC Model~\#2 from various combinations of \textcolor{blue}{acquired~(A)}, \textcolor{red}{harmonized~(H)}, and/or \textcolor{olive}{imputed~(I)} images as inputs, and segmented using THOMAS.
We compare with results from the acquired MPRAGE images segmented with THOMAS and HIPS-THOMAS.
Both the synthetic FGATIR images segmented with THOMAS and the acquired MPRAGE images segmented with HIPS-THOMAS yield statistically significantly higher DSC values than the acquired MPRAGE images segmented with THOMAS (Wilcoxon signed-rank test, p $<0.001$ for all comparisons, denoted by \(^{*}\)).
The best values in each column are indicated in \textbf{bold}.
}
\begin{tabular}{l cc cc cc cc cc}
\toprule
\textbf{Input Image} && \textbf{MPRAGE} && \textbf{T2w} && \textbf{FLAIR} && \textbf{Whole DSC} && \textbf{Mean DSC} \\
\cmidrule(lr){1-11}
\rowcolor{rowGreen}
Acq. MPRAGE (T) && -- && -- && -- && 0.7400 $\pm$ 0.252 && 0.6396 $\pm$ 0.220 \\
Acq. MPRAGE (HT) && -- && -- && -- && \textbf{0.8765 $\pm$ 0.017\(^*\)} && \textbf{0.8040 $\pm$ 0.031\(^*\)} \\
\rowcolor{rowGreen}
Synthetic FGATIR && \textcolor{blue}{A} && \textcolor{blue}{A} && \textcolor{olive}{I} && 0.8674 $\pm$ 0.033\(^*\) && 0.7713 $\pm$ 0.053\(^*\) \\
Synthetic FGATIR && \textcolor{blue}{A} && \textcolor{olive}{I} && \textcolor{olive}{I} && 0.8762 $\pm$ 0.027\(^*\) && 0.7808 $\pm$ 0.044\(^*\) \\
\rowcolor{rowGreen}
Synthetic FGATIR && \textcolor{red}{H} && \textcolor{red}{H} && \textcolor{olive}{I} && 0.8634 $\pm$ 0.032\(^*\) && 0.7715 $\pm$ 0.045\(^*\) \\
Synthetic FGATIR && \textcolor{red}{H} && \textcolor{olive}{I} && \textcolor{olive}{I} && 0.8628 $\pm$ 0.027\(^*\) && 0.7703 $\pm$ 0.040\(^*\) \\

\bottomrule
\\[-0.3em]
\end{tabular}
\label{table:mmti_thomas_dsc}
\end{table}

% \subsection{Synthesis for out-of-domain dataset}
\subsection{\Modified{Application to a real-world clinical trial dataset}{1(b)}{MODIFIED}}
\begin{figure}[!tb]
    \centering\includegraphics[width=0.8\linewidth]{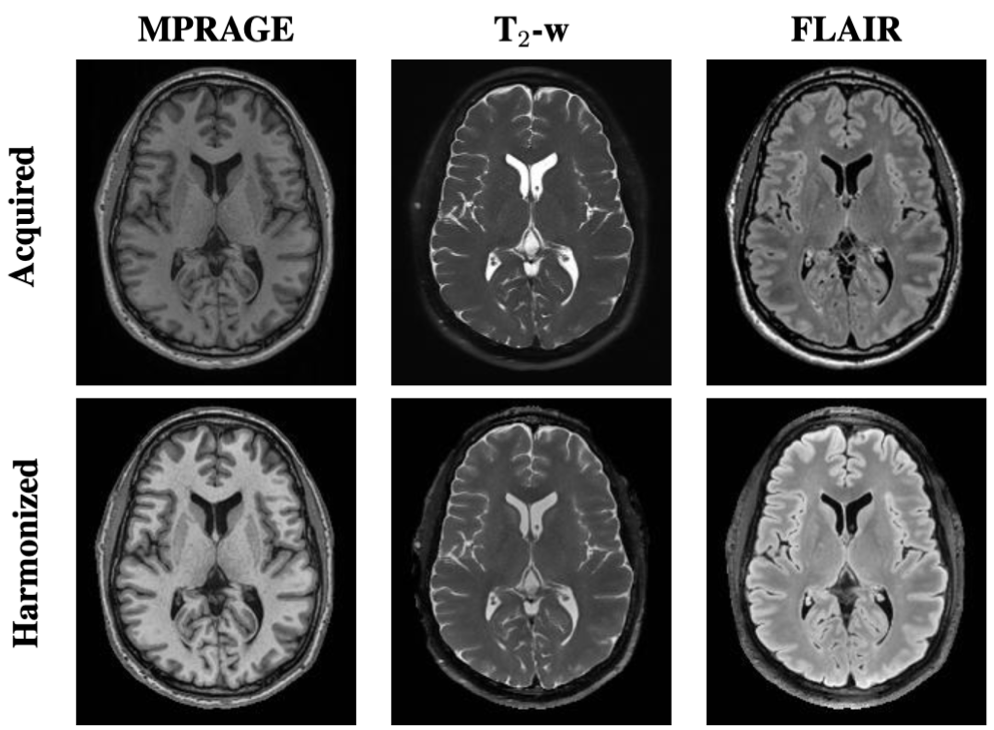}
    \caption{Images acquired in the PwMS dataset for one subject (top row) that were not acquired using the same protocol as our UMD training dataset. The harmonized images (bottom row), which have a similar contrast to our training dataset, are the input to our model.\label{fig:treat_acq_harm}
        }  %note label inside caption
\end{figure}

\begin{figure}[!tb]
    \centering\includegraphics[width=0.9\linewidth]{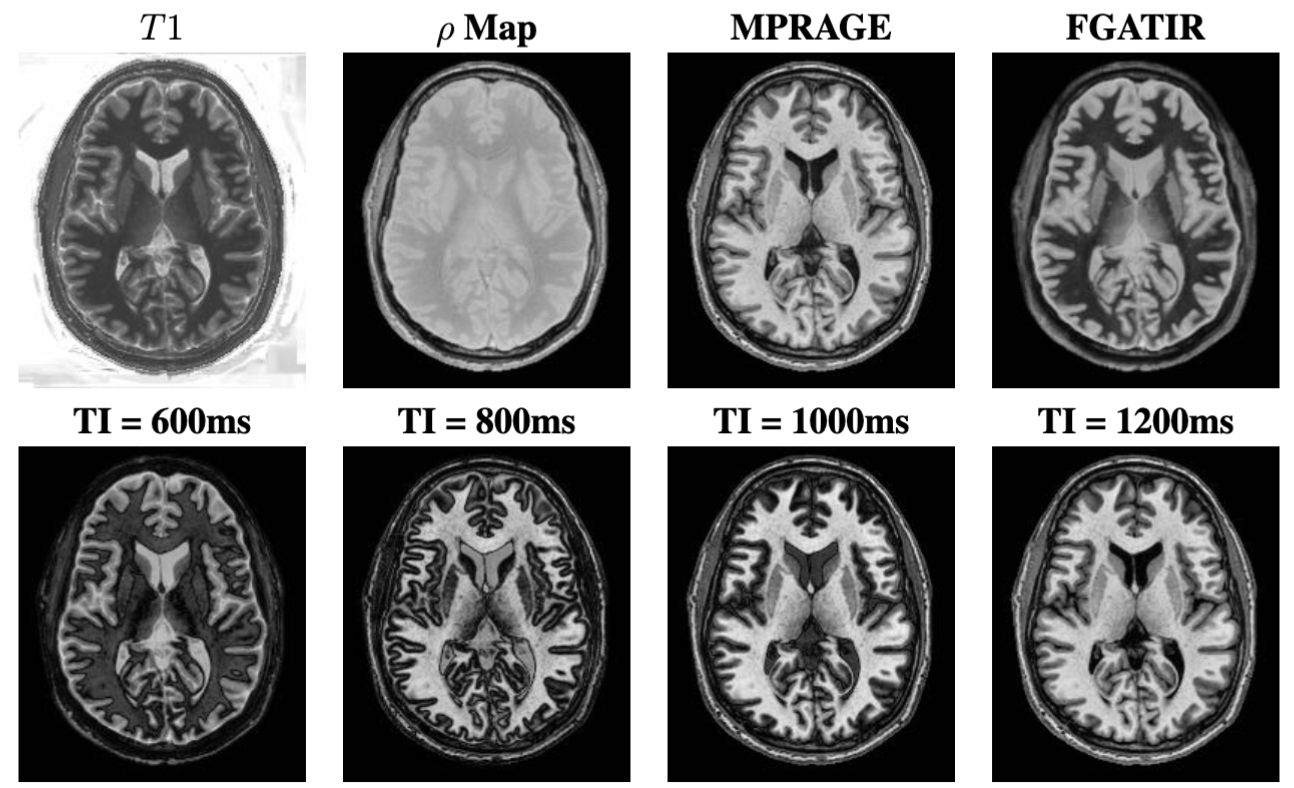}
    \caption{Predicted parameter maps and multi-TI images for an out-of-domain testing set using Model~\#2.}
\label{fig:treatms-multi}
\end{figure}

\Modified{Due to the scarcity of publicly available datasets with paired MPRAGE and FGATIR images, we applied our method to a real-world clinical trial dataset~\cite{mowry2025cct} of people with multiple sclerosis~(PwMS) to show additional qualitative results.}{1(b)}{MODIFIED}
When our testing dataset was not acquired with a similar acquisition protocol to our UMD training dataset, we used HACA3 for image harmonization and imputation.
An example of acquired and harmonized images from this out-of-domain dataset of PwMS is shown in Fig.~\ref{fig:treat_acq_harm}.
We used the harmonized images as input to Model~\#2.
The predicted parameter maps and calculated multi-TI images for one subject are shown in Fig.~\ref{fig:treatms-multi}.
Validation of this out-of-domain dataset is challenging due to the absence of ground truth FGATIR images. In the absence of direct FGATIR ground truth, we evaluated the fidelity of the synthesized MPRAGE images relative to that of the harmonized MPRAGE images.
The PSNR and SSIM of the synthetic MPRAGE images in the PwMS dataset are $33.76 \pm 2.28$ and $0.9967 \pm 0.0012$, respectively.
This experiment shows that by using HACA3 on out-of-domain data, SyMTIC is robust to changes in image acquisition parameters.

%\subsection{Ablations}
%Different loss functions and activations?

\section{Discussion and Conclusion}
In this work, we proposed SyMTIC, a model for multi-TI image synthesis using commonly acquired MR images: T$_1$-w, T$_2$-w, and FLAIR.
Our results demonstrate that SyMTIC can accurately synthesize multi-TI images from clinically acquired MR images by estimating the $T1$ and $\rho$ parameter maps.
While our approach requires three inputs,
we demonstrate how to handle situations where
not all inputs are available by using HACA3 for 
imputation. We also show HACA3’s ability to harmonize
out-of-domain data as a preprocessing step for our model.
The synthetic multi-TI images achieve accuracy comparable to those derived from acquired paired T$_1$-w images, yielding improved differentiation of thalamic nuclei and other deep brain structures, particularly for TI values in the range of 400–800 ms.

Our evaluation of different training losses revealed that using Model~\#2 for the SyMTIC model provides the best balance between parameter estimation and image synthesis accuracy.
We did not compare with direct synthesis using a U-Net because of its limited flexibility.
Training a direct synthesis model would require us to only synthesize a fixed set of TI images.
Our approach allows for customization of the TI value during inference.
The generation of arbitrary multi-TI images, in addition to FGATIR, offers potential advanced thalamic nuclei delineation, where different TIs enhance specific nuclei or subregions, enabling more accurate delineation.
Future work may explore alternative loss functions that explicitly incorporate tissue-specific constraints, ensuring that synthesized images retain clinically meaningful tissue contrast.

In this study, we used a 2D model rather than a 3D model to address computational and data limitations. Specifically, using 2D slices instead of full 3D volumes significantly reduced memory usage, allowing us to design a deeper network with sufficient capacity for learning. Additionally, this approach allowed us to generate approximately $300\times$ more training samples, as each 3D volume provided a large number of 2D slices from the three cardinal planes (sagittal, coronal, and axial). This increased the diversity of input samples, improving the robustness and generalizability of the model. With the acquisition of more data, future work should explore using a 3D model.

\Modified{One of the most significant outcomes of our study is showing how HACA3 can be used to help SyMTIC generalize to out-of-domain datasets with missing T2-w or FLAIR images along with unknown acquisition parameters. Other neuroimage processing algorithms, especially those trained on small datasets, might benefit from the incorporation of HACA3 as a pre-processing step.
Although, the reliance on HACA3 is a limitation.
If HACA3 introduces artifacts or fails to accurately map out-of-domain images into the in-domain space, it could impact the performance of SyMTIC.
To validate our reliance on HACA3, we used various combinations of imputation and harmonization as inputs to our SyMTIC model.
Thalamus segmentation results on the synthetic FGATIR images were similar to those achieved on the acquired MPRAGE using HIPS-THOMAS.
This showed SyMTIC can reliably synthesize images using a variety of synthetic images from HACA3.}{1(c)}{ADDED}

Additionally, our segmentation experiments using THOMAS indicate that synthetic FGATIR images provide superior thalamic segmentation accuracy compared to acquired MPRAGE images, with a mean DSC value of the entire thalamus improving from 0.820 (acquired MPRAGE) to 0.896 (synthetic FGATIR Model \#2).
This result is particularly important for applications such as DBS and MRgFUS, where accurate localization of deep gray matter structures is critical.
While our study confirms the benefits of synthetic FGATIR images for thalamic segmentation, further validation with expert-labeled datasets would strengthen confidence in its clinical applicability.

Another key aspect of our approach is that we retain the skull region in our synthetic images rather than applying skull-stripping, which is common in many MR preprocessing pipelines.
Although removal of the skull simplifies some analyses, retaining the skull provides a valuable anatomical context for applications where the interaction between the skull and brain is relevant~\cite{duan2022spie}.
For example, MRgFUS treatments rely on measurements of skull thickness and density to optimize ultrasound delivery, particularly in procedures for essential tremor and Parkinson’s disease.
By preserving the skull in our synthetic images, SyMTIC ensures compatibility with studies investigating these effects, while still enabling downstream analyses of deep gray matter structures.
Although we did not explicitly validate skull retention in our experiments, future work could explore its impact on MRgFUS targeting, skull-based registration techniques, and volumetric analyses in conditions affecting cranial morphology.

Our study also addresses the challenge of missing FLAIR images in clinical datasets.
We demonstrate that SyMTIC remains effective as evidenced by our experiments with the MMTI dataset, where the synthetic FLAIR image imputed using HACA3 successfully enabled multi-TI synthesis.
This capability is particularly relevant for retrospective studies, where incomplete imaging protocols limit the ability to perform advanced quantitative analysis.
Future work could explore the integration of using HACA3 to impute other missing MR contrasts such as T$_1$-w or T$_2$-w.

While our model focuses on $T1$ and $\rho$ parameter map synthesis, extending SyMTIC to estimate T$_2$ maps would significantly broaden its applicability.
The addition of T$_2$ mapping would enable direct calculation of a wider range of imaging contrasts beyond the inversion recovery multi-TI images discussed in this work.
Such an extension could further enhance applications in quantitative MRI, including relaxometry-based tissue characterization and lesion assessment in neurodegenerative diseases~\cite{jog2017random}.
A limitation of this study is that the synthetic model does not accurately model the physiological properties of non-brain tissues, such as fat and skull. While the focus was on generating complete images with anatomically plausible values, T1 representations for these regions may be inaccurate. Future work could aim to refine the modeling of non-brain tissues for applications requiring higher precision.

In summary, SyMTIC represents a novel approach for synthesizing multi-TI MR images from commonly acquired clinical sequences.
Our method enables enhanced visualization of deep gray matter structures, improves segmentation accuracy, and generalizes well to datasets with varying acquisition protocols.
By integrating harmonization and contrast imputation, SyMTIC offers a flexible solution for addressing the practical challenges of clinical imaging.
Future work should focus on refining the synthesis process, expanding parameter estimation to include additional tissue properties, and validating the method across larger multi-site datasets to further establish its clinical utility.

%\note{Note that the annoying labels which are showing can be removed by
%commenting out the line saying
%{\tt usepackage\{showkeys\}}.  But while
%drafting it is useful to have the labels/keys for the figures, equations
%etc showing.  Please turn them off when submitting the article for review. }

\section{Disclosures}
The authors declare that there are no financial interests, commercial affiliations, or other potential conflicts of interest that could have influenced the objectivity of this research or the writing of this paper.

\section{Code, Data, and Materials Availability}
All code associated with the model development and testing is publicly available at \url{https://github.com/UponAcceptance}.
The data utilized in this study were obtained from private datasets and not publicly available due to privacy concerns.

\section{Acknowledgments}
This work was partially supported by the National Science Foundation Graduate Research Fellowship under Grant No. DGE-2139757~(S.~P.~Hays) and National Cancer Institute (NCI) grants R01 CA253923~(L.~Zuo) and R01 CA275015~(L.~Zuo).
It was also supported by the PCORI grant MS-1610-37115 (PIs: S.~D.~Newsome and E.~M.~Mowry).
The statements in this publication are solely the responsibility of the authors and do not necessarily represent the views of PCORI, its Board of Governors or Methodology Committee.

\clearpage

%\section*{References}
% \addcontentsline{toc}{section}{\numberline{}References}
% \vspace*{-20mm}

\bibliography{cas-refs}
\bibliographystyle{spiejour}   % makes bibtex use spiejour.bst

%\vspace{1ex}
%\noindent Biographies and photographs of the other authors are not available.
\section{Biographies}

\textbf{Savannah~P.~Hays} is a Ph.D. candidate at Johns Hopkins University in the Department of Electrical and Computer Engineering.
Her research leverages artificial intelligence for MR image synthesis and harmonization. Savannah is a recipient of the prestigious National Science Foundation Graduate Research Fellowship (2023) and the Percy Pierre Fellowship from the Whiting School of Engineering (2022). She is an active member of the Baltimore-Washington section of the Society of Women Engineers.

\textbf{Lianrui~Zuo}, PhD, is a Postdoctoral Scholar at Vanderbilt University. He received his PhD at Johns Hopkins University in Electrical and Computer Engineering. His research lies at the intersection of medical imaging, machine learning, and clinical translation, with a focus on lung cancer risk modeling, image harmonization, and generative modeling. His work aims to advance reproducible science and improve clinical decision-making through innovative computational tools.

\textbf{Anqi~(Amy)~Feng} is a PhD student in Electrical and Computer Engineering at Johns Hopkins University. She is also affiliated with the National Institute on Aging, National Institutes of Health. Her research focuses on brain MRI analysis, with an emphasis on thalamic nuclei segmentation and multidimensional diffusion and relaxation MRI.

\textbf{Yihao Liu}, PhD, is a Research Assistant Professor in the Department of Electrical and Computer Engineering at Vanderbilt University. He received his Ph.D. in Electrical and Computer Engineering from Johns Hopkins University. His research focuses on medical image analysis, with particular expertise in deformable image registration, image harmonization, and retinal and neuroimaging applications.

\textbf{Blake E. Dewey}, PhD, is an Assistant Professor of Neurology at the Johns Hopkins University School of Medicine. He earned his Doctor of Philosophy in Electrical Engineering at the Johns Hopkins Whiting School of Engineering, where he focused on artificial intelligence solutions to data harmonization in MRI. His current research focuses on analyzing and modeling large clinical imaging datasets to advance translational research and precision medicine goals in neurodegeneration and neurological diseases.

\textbf{Jiachen Zhuo}, PhD, is an Associate Professor in the Department of Diagnostic Radiology and Nuclear Medicine at the University of Maryland. As Director of Clinical MR Physics and Co-Director of the MR Fellowship, she specializes in advanced neuroimaging techniques, MR pulse sequences, and MRgFUS. Her research focuses on traumatic brain injury and innovative applications of MRI technologies. Dr. Zhuo holds a PhD in Electrical and Computer Engineering from the University of Maryland, College Park.

\textbf{Ellen M. Mowry}, MD, is Professor of Neurology and Director of the Division of Neuroimmunology at Johns Hopkins University. She specializes in multiple sclerosis (MS), with research focusing on environmental and genetic factors influencing disease risk and progression, including vitamin D, diet, and the microbiome. A leading expert, she directs the MS Precision Medicine Center of Excellence and is a Principal Investigator on the Traditional Versus Early Aggressive Therapy for Multiple Sclerosis Trial.

\textbf{Scott D. Newsome}, DO, is Professor of Clinical Neurology at Johns Hopkins University, specializing in neuroimmunological and neuroinflammatory disorders such as multiple sclerosis, transverse myelitis, neuromyelitis optica, and stiff person syndrome. A leader in the field, Dr. Newsome is an adviser to the National MS Society and former president of the Consortium of MS Centers. He is currently a Principal Investigator on the Traditional Versus Early Aggressive Therapy for Multiple Sclerosis Trial.

\textbf{Jerry L. Prince}, the William B. Kouwenhoven Professor at Johns Hopkins University, is a leading expert in medical image analysis, including reconstruction, registration, and segmentation.
He has published over 500 papers relating to image processing and computer vision, and is a Fellow of IEEE, MICCAI, and AIMBE. 
Prince holds appointments across engineering, medicine, and data science, and has received numerous honors for his contributions to medical imaging.

\textbf{Aaron Carass}, a Research Scientist at Johns Hopkins University, has BA and MA degrees in mathematics from Trinity College Dublin, as well as MA, MEng, and DEng degrees in mathematics and computer science from Johns Hopkins University. He is a co-author on over 250 peer-reviewed articles, as well as co-authoring six book chapters, and has ten patents across a wide range of medical imaging modalities and applications.

\listoffigures
\listoftables

\end{spacing}
\end{document}